\documentclass[draft]{agujournal}
\usepackage{graphicx,epstopdf,titlesec,siunitx,amsmath,fixltx2e,textgreek,amsmath,lineno,makecell,tabularx,multirow,bigdelim}
\journalname{Geophysical Research Letters}

\begin{document}

%% ------------------------------------------------------------------------ %%
%  Title

\title{Ammonia in Jupiter's troposphere from high-resolution 5-\textmu m spectroscopy}

%% ------------------------------------------------------------------------ %%
%
%  AUTHORS AND AFFILIATIONS
%
%% ------------------------------------------------------------------------ %%

\authors{Rohini S. Giles\affil{1}, Leigh N. Fletcher\affil{2}, Patrick G. J. Irwin\affil{3}, Glenn S. Orton\affil{1} and James A. Sinclair\affil{1}}

\affiliation{1}{Jet Propulsion Laboratory / California Institute of Technology, 4800 Oak Grove Drive, Pasadena, CA 91109, USA}
\affiliation{2}{Department of Physics and Astronomy, University of Leicester, University Road, Leicester, LE1 7RH, UK}
\affiliation{3}{Atmospheric, Oceanic \& Planetary Physics, Department of Physics, University of Oxford, Clarendon Laboratory, Parks Road, Oxford, OX1 3PU, UK}

\correspondingauthor{R. S. Giles}{rohini.s.giles@jpl.nasa.gov}

\begin{keypoints}
\item Jupiter's tropospheric ammonia abundance is studied using high-resolution 5-\textmu m spectroscopy from VLT/CRIRES
\item The ammonia abundance decreases with altitude in the 1--4 bar pressure range at all latitudes
\item There is a strong localised enhancement in ammonia abundance at 4-6$^{\circ}$N
\end{keypoints}

%% ------------------------------------------------------------------------ %%

%% \begin{abstract} starts the second page 

\begin{abstract}
Jupiter's tropospheric ammonia (NH\textsubscript{3}) abundance is studied using spatially-resolved 5-\textmu m observations from CRIRES, a high resolution spectrometer at the European Southern Observatory's Very Large Telescope. The high resolving power (R=96,000) allows the line shapes of three NH\textsubscript{3} absorption features to be resolved. We find that within the 1--4 bar pressure range, the NH\textsubscript{3} abundance decreases with altitude. The instrument slit was aligned north-south along Jupiter's central meridian, allowing us to search for latitudinal variability. There is considerable uncertainty in the large-scale latitudinal variability, as the increase in cloud opacity in zones compared to belts can mask absorption features. However, we do find evidence for a strong NH\textsubscript{3} enhancement at 4--6$^{\circ}$N, consistent with a localised `ammonia plume' on the southern edge of Jupiter's North Equatorial Belt.
\end{abstract}

%% ------------------------------------------------------------------------ %%
%
%  TEXT
%
%% ------------------------------------------------------------------------ %%

%%%%%% INTRODUCTION %%%%%%

\section{Introduction}

As a condensible species in Jupiter's atmosphere, ammonia (NH\textsubscript{3}) plays an important role in understanding the planet's meteorology. At deep pressures in the atmosphere, the NH\textsubscript{3} abundance may be well-mixed, but the vertical profile becomes complex at higher altitudes. Assuming solar abundances, a typical cloud condensation model from~\citet{atreya99} suggests that NH\textsubscript{3} will start to dissolve in liquid water at $\sim$5.7 bar, and then will form two distinct cloud decks of solid NH\textsubscript{4}SH and solid NH\textsubscript{3} at 2.2 and 0.7 bar respectively. At even higher altitudes, gaseous NH\textsubscript{3} will be further depleted by photolysis due to UV solar radiation~\citep{atreya77}. Because the volume mixing ratio varies with altitude, NH\textsubscript{3} acts as a tracer for atmospheric dynamics and observations of the spatial distribution of NH\textsubscript{3} can be used to constrain global circulation patterns~\citep{showman05}.

NH\textsubscript{3} absorption features have been observed in jovian spectra across a wide range of wavelengths. Different spectral regions are sensitive to different altitudes in Jupiter's atmosphere and can be combined to constrain the vertical profile of NH\textsubscript{3}. In the infrared, there are NH\textsubscript{3} features at 45 \textmu m, 10 \textmu m and 5 \textmu m which probe the upper and middle parts of the troposphere~\citep[]{kunde82,irwin98,fouchet00b}. Additionally, NH\textsubscript{3} is a significant contributor to the microwave radiation from the planet; from the ground, this spectral region has been used to probe down to the 10-bar level in Jupiter's atmosphere~\citep{depater01} and the Microwave Radiometer (MWR) on the Juno mission is sensitive to the NH\textsubscript{3} abundance at pressures of up to 100 bar~\citep{li17}.

Previous 5-\textmu m studies of Jupiter's NH\textsubscript{3} abundance have been conducted using a range of data sources, including the Kuiper Airborne Observatory~\citep{bjoraker86a}, the Infrared Space Observatory~\citep{fouchet00b}, and spacecraft instruments such as Galileo NIMS~\citep{roos-serote98} and Juno JIRAM~\citep{grassi17}. While the previous microwave and long-wavelength infrared analyses of Jupiter's NH\textsubscript{3} abundance have explored spatial variability, these 5-\textmu m studies have typically either used disc-averaged spectra or focused on a specific area of interest, such as the 5-\textmu m hot-spots~\citep[e.g.][]{grassi17}. In this paper, we present latitudinally-resolved high-spectral resolution observations of Jupiter at 5 \textmu m, and we use these measurements to study the spatial variability in Jupiter's tropospheric NH\textsubscript{3} abundance. Sections~\ref{sec:observations} and~\ref{sec:spectral_modelling} describe the observations and our retrieval code. The results of the NH\textsubscript{3} retrievals are then presented in Section~\ref{sec:analysis} and discussed in Section~\ref{sec:discussion}.

%%%%%% OBSERVATIONS %%%%%%

\section{Observations}
\label{sec:observations}

\subsection{CRIRES observations}

The CRIRES instrument at the European Southern Observatory's Very Large Telescope (VLT) was used to make observations of Jupiter in the 5-\textmu m atmospheric window. CRIRES is a cryogenic high-resolution infrared echelle spectrograph~\citep{kaufl04}, which provides long-slit (0.2$\times$40'') spectroscopy with a resolving power of 96,000. The CRIRES observations used in this paper have previously been used to study the latitudinal variability in Jupiter's tropospheric disequilibrium species~\citep{giles17} and the H\textsubscript{3}\textsuperscript{+} auroral emission~\citep{giles16}. In this paper, these CRIRES observations are used to study NH\textsubscript{3} absorption in the 5-\textmu m window.

The observations used in this paper were made on 12 November 2012 at 05:30 UT. As described in~\citet{giles17}, the slit was aligned north-south along Jupiter's central meridian (at a longitude of 104$^{\circ}$W), allowing us to measure spatial variability with latitude, but not longitude. Observations were made in multiple wavelength settings across the 5-\textmu m spectral window, but this paper focuses on a single setting which covers the 5.15--5.19 \textmu m range and includes several strong NH\textsubscript{3} lines. In addition to observations of Jupiter, observations were made of a standard star, Pi-2 Orionis (HIP 22509), in order to provide radiometric calibration.

\subsection{Data reduction}

The data reduction process is described in detail in~\citet{giles17}. EsoRex, the ESO Recipe Execution Tool~\citep{ballester06}, was initially used to combine nodded pairs into a single observation and to achieve wavelength calibration. Subsequent data reduction was performed independently of the EsoRex software. The observations were straightened to correct observed distortions in both the spectral and spatial directions and were radiometrically calibrated using the observations of the standard star. Geometric information (including planetocentric latitude) was assigned to the observations using information from JPL HORIZONS.

%%%%%% SPECTRAL MODELLING %%%%%%

\section{Spectral modelling}
\label{sec:spectral_modelling}

The CRIRES observations were analysed using the NEMESIS retrieval algorithm~\citep{irwin08}. NEMESIS uses a multiple-scattering radiative transfer forward model to calculate the observed top-of-atmosphere spectral radiance for a given set of atmospheric parameters. These atmospheric parameters are then iteratively adjusted to match the observed spectrum, following an optimal estimation approach. As with~\citet{giles17}, line-by-line calculations are used in the forward model and this previous paper also details the sources of the spectroscopic line data. Again following the approach of~\citet{giles17}, we add a conservative 5\% forward modelling error to the retrieval calculations, to take into account inaccuracies in the model (e.g. line data). The error bars shown in Figures~\ref{fig:nh3_lines}, \ref{fig:nh3_eqz} and \ref{fig:nh3_segments} only show the noise on the observations, and any small discrepancies between the observations and the fitted spectra are likely due to these inaccuracies in the model. 

The reference atmospheric profile is described as in~\citet{giles15} and~\citet{giles17}. As with~\citet{giles17}, the atmosphere is divided into 39 pressure levels. The NH\textsubscript{3} abundance is allowed to vary continuously at each of the 39 pressure levels in order to determine its vertical profile. Although this paper focuses on NH\textsubscript{3} absorption features, H\textsubscript{2}O also has broad spectral features in the 5.15--5.19 \textmu m range, which contribute to the continuum radiance. In the retrievals presented in this paper, the H\textsubscript{2}O abundance was allowed to vary via a single relative humidity value. As a test, retrievals were also carried out where the H\textsubscript{2}O relative humidity was held fixed at values ranging from 1\% to 10\% (approximate range taken from~\citet{roos-serote00}); while this generally worsened the fit to the data, it did not affect the qualitative NH\textsubscript{3} results. 

In addition to gaseous species, tropospheric clouds also have a significant impact on the observed spectra at 5-\textmu m. As shown in many previous studies~\citep[e.g.][]{gierasch86,irwin01,giles15}, the banded appearance that Jupiter's exhibits at 5 \textmu m is due to the presence of clouds located between roughly 0.7 and 1.5 bar, which vary between optically thick in the zones and optically thin in the belts. \citet{giles17} showed that the CRIRES data could be modelled using a simple cloud model consisting of a single, spectrally-flat, compact cloud layer, located at 0.8 bar, with a single-scattering albedo $\omega=0.9$ and a Henyey-Greenstein asymmetry parameter $g=0.8$. This is used as the cloud model in this paper, and the cloud's optical thickness will be a free parameter in all retrievals. It should be noted that because the cloud particles are both highly scattering, and predominantly forward-scattering, the retrieved cloud optical thickness can be high, while still allowing photons to pass through. The potential impact of an additional deep cloud layer is discussed in Section~\ref{sec:beltzone_variability}.

%%%%%% NH3 ABSORPTION FEATURES %%%%%%
\section{Analysis}
\label{sec:analysis}

\subsection{NH\textsubscript{3} absorption features}
\label{sec:absorption_features}

\begin{figure*}
\centering
\makebox[\textwidth][c]{\includegraphics[width=1.15\textwidth]{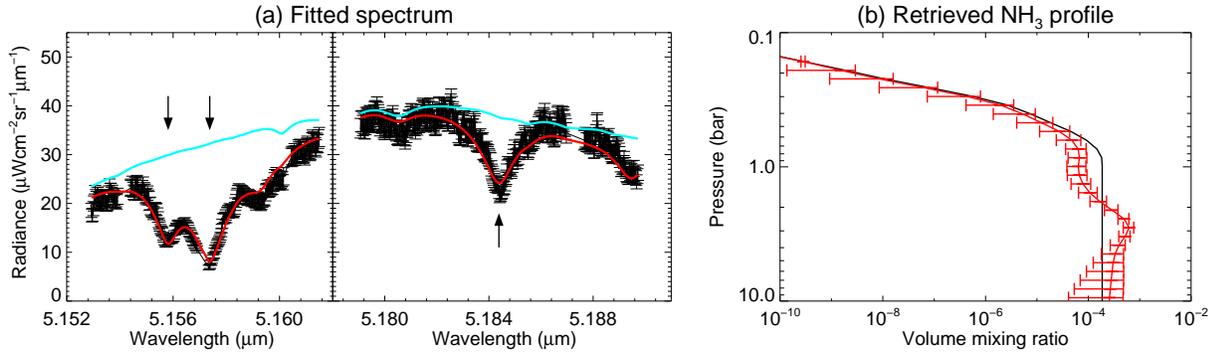}}
\caption{NH\textsubscript{3} absorption lines in the 5-\textmu m window from Jupiter's warm South Equatorial Belt (16$^{\circ}$S--6$^{\circ}$S). (a) The fitted spectrum (red) compared to the observational data (black). The blue line shows what the same spectrum would look like in the absence of NH\textsubscript{3}. The arrows show the NH\textsubscript{3} absorption line centres. (b) The retrieved NH\textsubscript{3} vertical profile (red) compared to the a priori profile (black).}
\label{fig:nh3_lines}
\end{figure*}

Jupiter's 5-\textmu m window contain several NH\textsubscript{3} absorption features. This paper focuses on three absorption lines which fall within the 5.15--5.19 \textmu m range. This spectral region was used because (i) it is relatively free of terrestrial absorption lines, which can be difficult to remove cleanly; and (ii) it is relatively free from strong absorption features from other jovian tropospheric species. The locations of the three  NH\textsubscript{3} absorption lines are shown by the arrows in Figure~\ref{fig:nh3_lines}(a). The lines at 5.156 and 5.157 \textmu m are relatively strong, while the line at 5.184 \textmu m is relatively weak. Depending on their strength, different lines probe slightly different pressure levels in Jupiter's atmosphere. Assuming the a priori atmospheric profile and a cloud-free atmosphere, the three absorption lines have a maximum sensitivity at 1.64, 1.56 and 3.19 bar respectively. By simultaneously fitting all three lines, we can therefore constrain the vertical profile of NH\textsubscript{3} within the $\sim$1--4 bar pressure range. 

The black lines in Figure~\ref{fig:nh3_lines}(a) show the CRIRES observations of Jupiter. This data is taken from the planet's warm, cloud-free South Equatorial Belt (SEB, 16$^{\circ}$S--6$^{\circ}$S); this part of the planet is bright at 5 \textmu m, so the noise on the data (shown by the error bars) is relatively low. NEMESIS was used to perform a retrieval on this data, and the resulting fitted spectrum is shown by the red line plotted on top of the data. For comparison, the blue line shows what the same fitted spectrum would look like if NH\textsubscript{3} were not present in Jupiter's atmosphere. 

In order to fit the CRIRES data, the NH\textsubscript{3} abundance was allowed to vary at each of the 39 pressure levels in the model atmosphere. The retrieved vertical profile is shown in red in Figure~\ref{fig:nh3_lines}(b), alongside the a priori profile in black, taken from analyses of Cassini/CIRS data~\citep{fletcher09}. As expected, the region of maximum sensitivity is at $\sim$1--4 bar; outside this range, the retrieval tends towards the a priori due to a lack of other information and the error bars tend towards the 100\% errors placed on the a priori. 

Although the a priori vertical profile from~\citet{fletcher09} assumes a constant NH\textsubscript{3} abundance at all pressures greater than 0.8 bar, the retrieved vertical profile in Figure~\ref{fig:nh3_lines}(b) deviates from this assumption. Instead, a higher abundance is required at higher pressures (primarily driven by the 5.184 \textmu m line) and a lower abundance is required at lower pressures (primarily driven by the 5.156 and 5.157 \textmu m lines). This agrees with previous results from~\citet{fouchet00b} and Section~\ref{sec:latitudinal_retrieval} shows that this pattern extends across the planet.

%%%%%% LATITUDINAL RETRIEVAL %%%%%%

\subsection{Latitudinal variability}
\label{sec:latitudinal_retrieval}

\begin{figure}
\centering
\includegraphics[width=9.5cm]{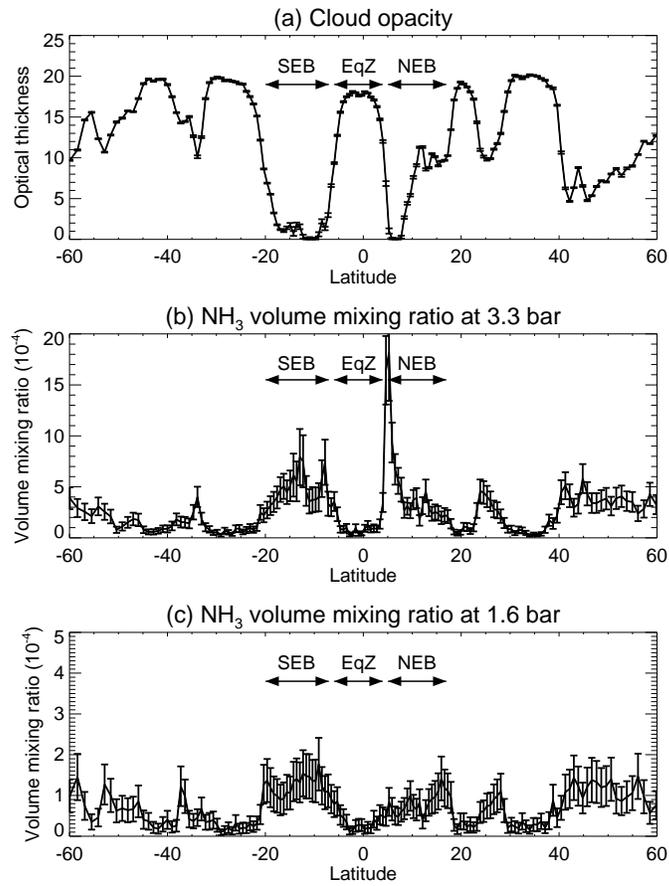}
\caption{Latitudinal retrieval of CRIRES observations. The retrieved cloud opacity as a function of latitude is shown in (a). The retrieved NH\textsubscript{3} volume mixing ratios at 3.3 and 1.6 bar are are shown in (b) and (c) respectively.}
\label{fig:nh3_latitude}
\end{figure}

Section~\ref{sec:absorption_features} focused on a single spatial region of the planet. In this section, we will now explore how the NH\textsubscript{3} vertical profile varies with latitude across Jupiter. As described in Section~\ref{sec:observations}, the CRIRES observations provide latitudinally resolved spectra from Jupiter's central meridian. The spectra were smoothed in the latitudinal direction with a width of 5 pixels in order to reduce the noise, and the sampling rate was 3 pixels. For each smoothed spectrum, a retrieval was performed using NEMESIS where the NH\textsubscript{3} vertical profile was allowed to vary continuously. As in Figure~\ref{fig:nh3_lines}, the retrieval assumes a single cloud layer with a variable optical thickness. The results are shown in Figure~\ref{fig:nh3_latitude}; (a) shows the retrieved cloud opacity as a function of latitude, while (b) and (c) show the retrieved NH\textsubscript{3} volume mixing ratios at 3.3 and 1.6 bar respectively.  The previous section showed that the SEB retrievals lead to a volume mixing ratio that decrease with altitude in the 1--4 bar pressure range. Figure~\ref{fig:nh3_latitude} shows that this remains true at all latitudes, as (c) shows consistently lower abundances than (b).

As a comparison, retrievals were also performed where NH\textsubscript{3} was assumed to be well-mixed at high pressures, before dropping to zero at 0.8 bar (the approximate condensation point of NH\textsubscript{3}). This led to a consistently poorer fit to the data, but to varying degrees across the planet. In the cool zones, this change did not have a large impact, increasing the goodness-of-fit parameter (\textchi\textsuperscript{2}) by \textless5\%. In the warm belts, the well-mixed vertical profile increased \textchi\textsuperscript{2} by $\sim$20\%. The greatest impact was from spectra near 5$^{\circ}$N, where a vertical profile that decreases with altitude fits the data 100\% better than a well-mixed profile.

\subsubsection{Belt-zone variability}
\label{sec:beltzone_variability}

Figure~\ref{fig:nh3_latitude}(a) shows the belt-zone structure of Jupiter's atmosphere. In the planet's zones (such as the Equatorial Zone), the clouds are optically thick. In the adjacent belts, the cloud opacity drops significantly. Figures~\ref{fig:nh3_latitude}(b) and (c) show that some of the retrieved variability in the NH\textsubscript{3} abundance coincides with the belt-zone structure, with higher retrieved abundances in the belts. In particular, the increase in NH\textsubscript{3} abundance that can be seen at 5--20$^{\circ}$S in both (b) and (c) coincides with the SEB. While it is possible that these retrievals represent a genuine increase in NH\textsubscript{3} at these latitudes, it is also possible that this is an artefact due to the tropospheric cloud structure.~\citet{bjoraker15} and~\citet{giles17} showed that there is evidence for a $\sim$5-bar water cloud that varies between optically thick in the zones and optically thin in the belts.~\citet{giles17} also showed that the presence of this cloud can explain similar apparent belt-zone variability in the line shapes of tropospheric gaseous species.

\begin{figure*}
\centering
\makebox[\textwidth][c]{\includegraphics[width=1.15\textwidth]{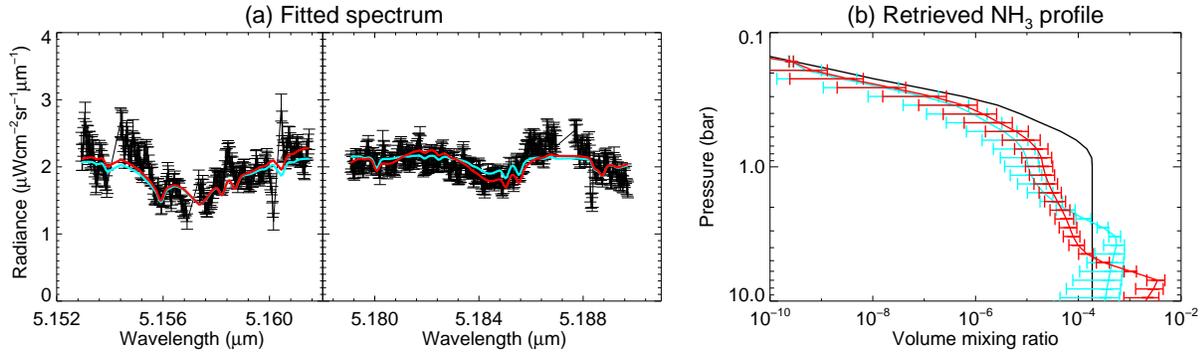}}
\caption{NH\textsubscript{3} absorption lines in the 5-\textmu m window from Jupiter's cool Equatorial Zone (5$^{\circ}$S--4$^{\circ}$N). (a) The fitted spectrum assuming that there is no deep cloud (red) compared to the fitted spectrum assuming that there is an opaque cloud at 5 bar (blue). (b) The retrieved NH\textsubscript{3} vertical profiles in the two cases: no deep cloud (red) and opaque 5-bar cloud (blue).}
\label{fig:nh3_eqz}
\end{figure*}

This is shown further by Figure~\ref{fig:nh3_eqz}, which shows the NH\textsubscript{3} absorption features from the Equatorial Zone (EqZ, 5$^{\circ}$S--4$^{\circ}$N). When compared to Figure~\ref{fig:nh3_lines}, it is clear that the absorption features are much shallower in the EqZ than in the SEB. One interpretation of this is that the NH\textsubscript{3} abundance is higher in the SEB and lower in the EqZ. The red lines in Figure~\ref{fig:nh3_eqz} show the fitted spectrum and retrieved profile if there is no deep cloud present in the atmosphere, and they lead to this conclusion. Over the sensitive 1--4 bar region, the retrieved abundances shown in red in Figure~\ref{fig:nh3_eqz}(b) are significantly lower than those in Figure~\ref{fig:nh3_lines}(b). However, this trend can be cancelled out, or even reversed, by the addition of a deep cloud layer in the EqZ. Following~\citet{giles17}, an opaque cloud was inserted at 5 bar. The results of this retrieval are shown by the blue lines in Figure~\ref{fig:nh3_eqz}. While Figure~\ref{fig:nh3_eqz}(a) shows that the fit remains similar, the retrieved abundances in Figure~\ref{fig:nh3_eqz}(b) are fairly different. Because the presence of the deep cloud suppresses the absorption features, the retrieved abundances in the 1--4 bar region are now higher. In fact, the 3-bar abundance is now slightly higher for the EqZ than for the SEB. The deep cloud structure therefore leads to considerably uncertainty in the belt-zone variability in NH\textsubscript{3}. 

\subsubsection{Small-scale variability}

While deep tropospheric clouds can plausibly explain the retrieved NH\textsubscript{3} enhancement in the SEB, the same is not true for the observed spike at 5$^{\circ}$N shown in Figure~\ref{fig:nh3_latitude}(b). This spike is offset from the warm, cloud-free North Equatorial Belt (NEB). While NH\textsubscript{3} abundance peaks at 5$^{\circ}$N, the minimum cloud opacity is at 7$^{\circ}$N, by which point the NH\textsubscript{3} abundance has dropped to a third of its maximum value. While Figure~\ref{fig:nh3_latitude}(a) shows the opacity of the 0.8-bar cloud,~\citet{giles17} demonstrated that the deep cloud structure follows the same latitudinal pattern as the upper cloud, reaching a minimum in the centre of the belts. Based on the cloud structure alone, we would expect the maximum retrieved NH\textsubscript{3} abundance to be at 7$^{\circ}$N. It therefore appears that there is a genuine enhancement of NH\textsubscript{3} at 4--6$^{\circ}$N and a corresponding relative depletion at 7-9$^{\circ}$N at 3.3 bar. 

\begin{figure*}
\centering
\makebox[\textwidth][c]{\includegraphics[width=1.15\textwidth]{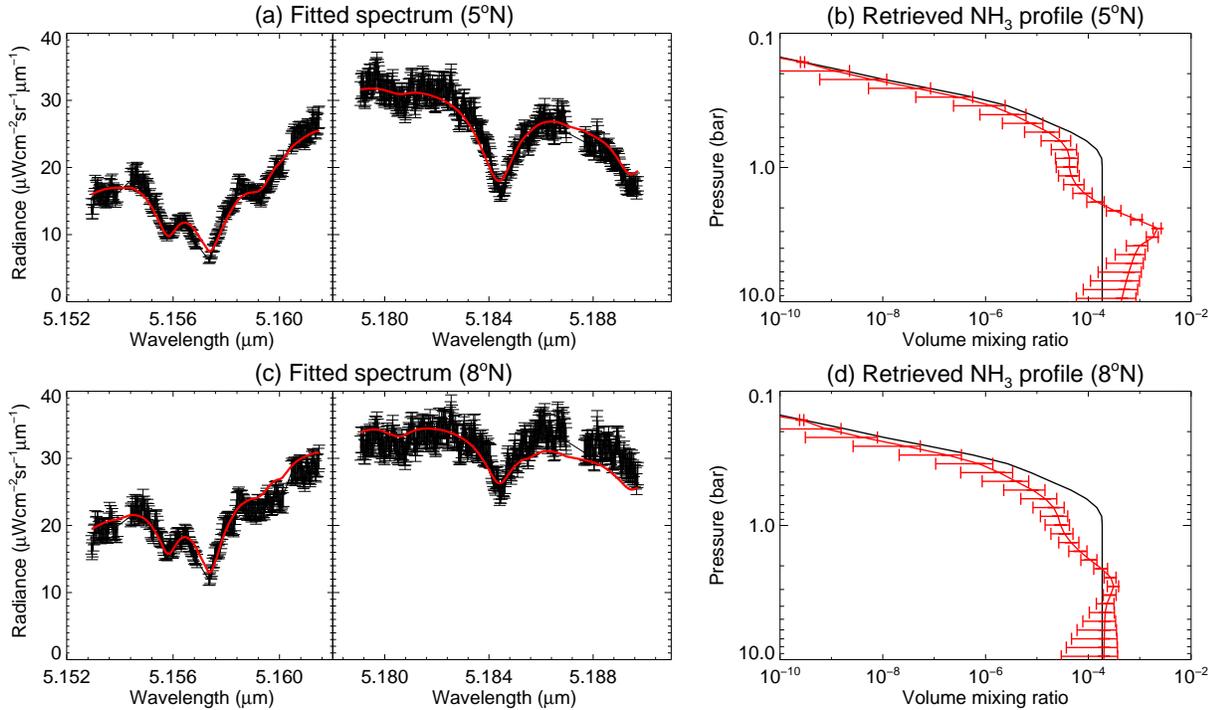}}
\caption{NH\textsubscript{3} absorption features from two different points on the planet: 5$^{\circ}$N and 8$^{\circ}$N. (a) and (c) show the fitted spectrum (red) compared to the observational data (black). (b) and (d) show the retrieved NH\textsubscript{3} vertical profile (red) compared to the a priori profile (black).}
\label{fig:nh3_segments}
\end{figure*}

This is further demonstrated in Figure~\ref{fig:nh3_segments} which shows the results of the retrievals from these two points on the planet. Figures~\ref{fig:nh3_segments}(a) and~\ref{fig:nh3_segments}(c) show the observed spectra from 5$^{\circ}$N and 8$^{\circ}$N, along with the retrieved best-fit spectrum (with a single 0.8-bar cloud layer). The absorption features at  5$^{\circ}$N are deeper than the absorption features at 8$^{\circ}$N. The average radiance in both cases is very similar; this is primarily determined by the cloud opacity, and therefore shows that these two regions have similar amounts of cloud cover. The differences between them are therefore primarily due to differences in the NH\textsubscript{3} abundance. This is further emphasised by the fact that if there were any differences in the deep water cloud opacity, the opacity is likely to be higher at 5$^{\circ}$N than at 8$^{\circ}$N, since the former is closer to the cloudy Equatorial Zone. However, thick deep clouds have a tendency to obscure absorption features, rather than enhance them, so they cannot explain the spectral differences between these two cases.

The most striking difference between Figures~\ref{fig:nh3_segments}(a) and~\ref{fig:nh3_segments}(c) is in the depth of the absorption feature at 5.184 \textmu m, which probes higher pressures in the atmosphere; it is significantly deeper at 5$^{\circ}$N than at 8$^{\circ}$N. In contrast, the features at 5.156 and 5.157 \textmu m, which probe lower pressures, are relatively similar in the two cases. The retrieved NH\textsubscript{3} profiles (Figure~\ref{fig:nh3_segments}(b) and~\ref{fig:nh3_segments}(d)) reflect these spectral similarities and differences; at higher pressures ($\sim$3 bar) there is a large difference between the retrieved NH\textsubscript{3} abundances, but at lower pressures ($\sim$1 bar) the retrieved abundances are approximately the same. This is also reflected in Figure~\ref{fig:nh3_latitude}, which shows a 5$^{\circ}$N enhancement, but a relatively flat distribution at 1.6 bar.

%%%%%%%%%%%%%%%%%%%%%

\section{Discussion and conclusions}
\label{sec:discussion}

High-resolution ground-based observations from the CRIRES instrument at the VLT were used to analyse NH\textsubscript{3} absorption features in Jupiter's 5-\textmu m spectrum. The three absorption features at 5.15--5.19 \textmu m probe the 1--4 bar pressure range in Jupiter's troposphere, allowing the vertical profile to be constrained. The CRIRES observations show that the NH\textsubscript{3} abundance decreases with altitude over this pressure range. This is consistent with the results of a previous 5-\textmu m study by~\citet{fouchet00b}, who found that the NH\textsubscript{3} mixing ratio increases downward to at least the 4-bar level. This agrees with the nominal cloud structure of the planet~\citep{atreya99}, in which NH\textsubscript{3} forms cloud layers of aqueous-ammonia solution, ammonium hydrosulphide and ammonia ice at 5.7, 2.2 and 0.7 bar respectively.

The CRIRES observations were then used to study the latitudinal variability in the NH\textsubscript{3} abundance. Using a simple one-cloud model, there is some belt-zone variability in the retrieved abundances, with higher volume mixing ratios observed in the cloud-free belts and lower volume mixing ratios observed in the cloudy zones. However, this trend can be removed by adding in an optically-thick 5-bar cloud in the zones, as previously shown by~\citet{giles17}. If this cloud is made opaque, the retrieved 3-bar NH\textsubscript{3} abundance is actually slightly higher in the EqZ than in the SEB. The effects of the deep cloud add considerably uncertainty comparisons of spectra from the belts and the zones.  

However, spectra from regions with similar cloud cover can be robustly compared, and the CRIRES data provide evidence for smaller-scale spatial variability in NH\textsubscript{3}. We find an NH\textsubscript{3} enhancement at 4--6$^{\circ}$N and a corresponding relative depletion at 7--9$^{\circ}$N, at a pressure of 3--4 bar, despite the fact that these two spatial regions have similar cloud opacities. In addition, we find that this enhancement does not continue through to higher altitudes; by 1.6 bar, the trend has disappeared.

This is consistent with the `ammonia plumes' detected by~\citet{fletcher16} using 10-\textmu m observations from the TEXES instrument at NASA's Infrared Telescope Facility. TEXES was used to make global maps of the planet, and the spectra are sensitive to the NH\textsubscript{3} abundance at the 500-mbar level. \citet{fletcher16} found a series of NH\textsubscript{3}-rich plumes at latitudes of around 5$^{\circ}$N, located directly to the south-east of NH\textsubscript{3}-dessicated hot-spots. At 3--4 bar, the CRIRES observations show the same latitudinal profile, albeit limited to a single longitude. However, it is unclear why the plume would be evident at \textgreater3 bar and 500 mbar, but be absent at 1.6 bar. 

While the TEXES observations probed higher altitudes than CRIRES, observations were recently made with the Juno MWR instrument that are sensitive to the NH\textsubscript{3} abundance at pressures ranging from 500 mbar up to 100 bar~\citep{janssen17}.~\citet{li17} found that NH\textsubscript{3} had an asymmetrical distribution, with an enhancement at 0--5$^{\circ}$N and a depletion at 5--15$^{\circ}$N. Like CRIRES, MWR shows a sharp contrast in NH\textsubscript{3} abundance between 4--6$^{\circ}$N and 7--9$^{\circ}$N. However the strong equatorial enhancement seen by MWR is not observed in the CRIRES data. This can be partially explained by the presence of deep, thick clouds in the Equatorial Zone, which act to suppress the gaseous absorption features at 5-\textmu m and can therefore increase the retrieved abundances if they are included in the model. However, even an entirely opaque deep cloud cannot reproduce the strong enrichment at all pressure levels that is seen at the equator in the MWR observations. The cause of the discrepancy between the two spectral regimes is unclear and should be the focus of future studies; possible explanations could include temporal variability (between 2012 and 2016), temperature variations, or additional complex cloud effects.

\acknowledgments

This work is based on observations collected at the European Organisation for Astronomical Research in the Southern Hemisphere under ESO programme 090.C-0053(A). The data is available from the ESO Science Archive Facility. The research was carried out in part at the Jet Propulsion Laboratory, California Institute of Technology, under a contract with the National Aeronautics and Space Administration. Giles and Sinclair were supported by the NASA Postdoctoral Program, and Orton was supported by grants from NASA to the Jet Propulsion Laboratory/California Institute of Technology. Fletcher was supported by a Royal Society Fellowship at the University of Leicester and Irwin was supported by the UK Science and Technology Facilities Council. 

\textcopyright~2017. All rights reserved

\bibliography{/Users/rohinigiles/Documents/Work/master.bib}

\end{document}